\documentclass[%
 reprint,
 amsmath,amssymb,
 aps,
 pra,
]{revtex4-2}

\usepackage{graphicx}
\usepackage{pgfplots}
\pgfplotsset{compat=1.18}

\usepackage{dcolumn}
\usepackage{bm}
\usepackage{hyperref}
\hypersetup{
    colorlinks,
    linkcolor={black},
    citecolor={black},
    urlcolor={blue}
}
\usepackage{tikz}
\usetikzlibrary{decorations.pathreplacing}
\usetikzlibrary{arrows.meta}
\usetikzlibrary{decorations.pathmorphing}
\usetikzlibrary{positioning}
\usetikzlibrary{calc}

\usepackage{adjustbox}
\usepackage{makecell}

\usepackage{xfrac}

\usetikzlibrary{decorations.pathmorphing}

\usepackage{booktabs}

\newcommand{\cm}{\,\mathrm{cm}}

\newcommand{\kHz}{\,\mathrm{kHz}}
\newcommand{\MHz}{\,\mathrm{MHz}}
\newcommand{\GHz}{\,\mathrm{GHz}}
\newcommand{\THz}{\,\mathrm{THz}}

\newcommand{\mm}{\,\mathrm{mm}}
\newcommand{\nm}{\,\mathrm{nm}}
\newcommand{\mW}{\,\mathrm{mW}}

\newcommand{\vms}{\,\mathrm{m/s}}
\newcommand{\us}{\,\mu\mathrm{s}}
\newcommand{\mWcms}{\,\mathrm{mW/}\mathrm{cm}^2}

\newcommand{\CENTREX}{CeNTREX}

\newcommand{\RzeroFone}{$R(0)\,\tilde{F}^\prime_1=1/2\,F^\prime=1$}
\newcommand{\RzeroFtwo}{$R(0)\,\tilde{F}^\prime_1=3/2\,F^\prime=2$}
\newcommand{\RoneFthree}{$R(1)\,\tilde{F}^\prime_1=5/2\,F^\prime=3$}
\newcommand{\RtwoFfour}{$R(2)\,\tilde{F}^\prime_1=7/2\,F^\prime=4$}
\newcommand{\RthreeFfive}{$R(3)\,\tilde{F}^\prime_1=9/2\,F^\prime=5$}
\newcommand{\PtwoFone}{$P(2)\,\tilde{F}^\prime_1=3/2\,F^\prime=1$}

\newcommand{\bra}[1]{\langle#1|}
\newcommand{\ket}[1]{|#1\rangle}
\newcommand{\braket}[2]{\langle#1|#2\rangle}

\newcommand{\parts}[1]{\textit{\textbf{#1.}}}

\newcommand{\brzero}{bf_{F=0}}
\newcommand{\brone}{bf_{F=1}}
\newcommand{\bronezero}{bf_{1/0}}

\newcommand{\tlfground}{$\mathrm{X}\,^1\Sigma^+$}
\newcommand{\tlfgroundmath}{\mathrm{X}\,^1\Sigma^+}
\newcommand{\tlfexc}{$\mathrm{B}\,^3\Pi_1$}
\newcommand{\tlfexcmath}{\mathrm{B}\,^3\Pi_1}
\newcommand{\maxgain}{29.4(3)}

\begin{document}

\preprint{APS/123-QED}

\title{Rotational-hyperfine cooling of $^{205}$TlF in a cryogenic beam}

\author{Olivier Grasdijk}
 \email{jgrasdijk@argonne.gov}
\author{David DeMille}%
 \altaffiliation[]{Department of Physics and Astronomy, Johns Hopkins University, Baltimore, MD 21218, United States of America}{}
\altaffiliation[]{Department of Physics, University of Chicago, Chicago, IL 60637, United States of America}{}
 
\affiliation{%
 Physics Division, Argonne National Laboratory, Argonne, IL 60439, United
States of America
}%

\author{Jakob Kastelic}
\author{Steve Lamoreaux}
\author{Oskari Timgren}
\affiliation{%
 Department of Physics, Yale University, New Haven, CT 06511, United States of America
}%

\author{Konrad Wenz}
\author{Tanya Zelevinsky}
\affiliation{
 Department of Physics, Columbia University, New York, NY 10027-5255, United States of America
}
\author{David Kawall}
\affiliation{
 Department of Physics, University of Massachusetts Amherst, Amherst, MA
 01003, United States of America
}

\collaboration{CeNTREX Collaboration}

\date{\today}

\begin{abstract}
The aim of \CENTREX\ (Cold Molecule Nuclear Time-Reversal Experiment) is to search for time-reversal symmetry violation in the thallium nucleus, by measuring the Schiff moment of $^{205}$Tl in the polar molecule thallium fluoride (TlF). CeNTREX uses a cryogenic beam of TlF with a rotational temperature of 6.3(2) K. This results in population spread over dozens of rotational and hyperfine sublevels of TlF, while only a single level is useful for the Schiff moment measurement. Here we present a protocol for cooling the rotational and hyperfine degrees of freedom in the CeNTREX beam, transferring the majority of the Boltzmann distribution into a single rotational and hyperfine sublevel by using a single ultraviolet laser and a pair of microwave beams. We achieve a factor of $20.1(4)$ gain in the population of the $J=0$, $F=0$ hyperfine sublevel of the TlF ground state. 
\end{abstract}

 
\maketitle


\section{Introduction}
\label{sec:introduction}

\CENTREX\ (Cold Molecule Nuclear Time-Reversal Experiment) aims to achieve a significant increase in sensitivity over the best present upper bounds on the strength of certain hadronic time-reversal ($T$) violating fundamental interactions such as the proton's electric dipole moment and the $CP$-violating parameter of quantum chromodynamics, $\theta_{\rm QCD}$. 
The overall approach of CeNTREX, and details of its measurement strategy, are described in~\cite{Grasdijk_2021}. \CENTREX\ seeks to determine the Schiff moment of the $^{205}$Tl nucleus by performing magnetic resonance measurements on the nucleus within an electrically polarized thallium monofluoride (TlF) molecule. The strongly polarized electron shells in the molecule interact strongly with the Schiff moment, providing orders of magnitude larger shifts of the magnetic resonance frequency than in experiments using atoms, for the same size of Schiff moment~\cite{sandars1967measurability}.

CeNTREX makes use of a cryogenic buffer gas cooled molecular beam source \cite{Grasdijk_2021,hutzler2012buffer} to create a beam of TlF. This source delivers a beam with high intensity (for good statistics), low mean velocity (to enable long interaction time and good energy resolution), low velocity dispersion (to enable efficient electrostatic focusing \cite{Wu_2022}) and low rotational temperature (to reduce the spread of population over many internal states). The CeNTREX source uses a neon buffer gas at a temperature of $19$~K. Compared to colder sources using helium buffer gas, this allows operation at higher repetition rates and gives steadier flux of very heavy species \cite{C1CP20901A} such as TlF.  

After some cooling during the free expansion of the neon gas as it exits the source, the TlF has a rotational temperature $T_\mathrm{rot} = 6.3(2)$\,K~\cite{Grasdijk_2021}. At this temperature nearly all molecules are assumed to be in the vibronic ground state. Only $\sim\!5 \%$ of the TlF molecules are in the lowest rotational level $J=0$ (where $J$ is the rotational quantum number), and only $1/4$ of these are in the absolute hyperfine/rotational ground state $J=0,~F=0$ [where $F$ is the total hyperfine angular momentum including rotation and the nuclear spins of $^{205}$Tl ($I_1 = 1/2$) and $^{19}$F ($I_2 = 1/2$)]. Only this single sublevel is used for the Schiff moment measurement protocol in CeNTREX \cite{Grasdijk_2021}. 

Here we describe a method to dramatically increase the population of this level, and hence to improve the statistical sensitivity of the experiment. 
We refer to our method as rotational/hyperfine cooling of the TlF molecules. While rotational cooling has been previously performed in molecules and molecular ions \cite{PhysRevLett.115.233001, Courageux_2022}, it has not been done for hyperfine structure or a closed-shell molecule before. Using a combination of laser optical pumping and microwave-driven rotational state transfer, we drive most of the population from all hyperfine sublevels of the three lowest excited rotational levels ($J=1,2,$ and $3$) into the $J=0$ manifold of states, preferentially into the $F=0$ sublevel.  In the Boltzmann distribution at $T_\mathrm{rot}=6.3(2)$\,K, approximately $56\%$ of the population is in the lowest four rotational states, i.e. $J=0-3$.  If all population in the $J=0,1,2$, and $3$ rotational ground states were transferred to the $\ket{J=0, F=0}$ state, a maximum gain of $\approx\! 45$ in its population could be achieved. Though our method incorporates a simplification that reduces the maximum potential gain to $\approx\! 29$, this enhancement in signal size for the $^{205}$Tl Schiff moment measurement is crucial for CeNTREX to attain its projected sensitivity \cite{Grasdijk_2021}.

In what follows, we explain the principle of our method for rotational/hyperfine cooling of TlF, and present experimental results on its implementation.

\section{Principle of the Method}
\label{sec:principle}

To explain our method, we first review relevant aspects of the structure of TlF (see e.g.~\cite{Grasdijk_2021} for more details). Throughout, unless otherwise noted, we discuss only states with vibrational quantum number $v=0$. The ground state \tlfground\, has a rotational constant $B = 6.66733$~GHz and nominal rotational energies $E_J=BJ(J+1)$. The $^{205}$Tl and $^{19}$F nuclear spins give rise to hyperfine substructure in the rotational states. The lowest rotational state, $\ket{J=0}$, splits into two levels, with $F=0$ and $F=1$. All higher rotational states split into four hyperfine states, separated over two $\mathbf{F_1} = \mathbf{J} + \mathbf{I_1}$ hyperfine manifolds, with the $F_1 = J - I_1$ taking $F=J-1$ and $J$, and $F_1 = J+I_1$ taking $F=J$ and $J+1$. Hyperfine splittings in the low-$J$ levels of interest here are always less than 600~kHz.  In the $J=0$ state, $F=1$ is split from $F=0$ by only 13~kHz.
Transitions between states with $\Delta J = \pm 1$ can be coupled with microwave electromagnetic fields. 
\begin{figure*}
	\centering
	\small
	\begin{minipage}[c]{0.49\textwidth}
		\begin{tikzpicture}

    \node (J0) at (0,0) {$\ket{J=0^+}$};
    \node (J1) at (2,.5) {$\ket{J=1^-}$};
    \node (J2) at (4,1) {$\ket{J=2^+}$};
    \node (J3) at (6,1.5) {$\ket{J=3^-}$};
    \node (Je1) at (3,4) {$\ket{\widetilde{J'}=1^-,\widetilde{F^\prime_1}=3/2,F^\prime=1}$};

    \node[below=-5pt of J0] (p0) {\footnotesize5.0~\%};
    \node[below=-5pt of J1] (p1) {\footnotesize13.4~\%};
    \node[below=-5pt of J2] (p2) {\footnotesize18.3~\%};
    \node[below=-5pt of J3] (p3) {\footnotesize19.0~\%};

    \node (GS) at (7.2,.75) {$X^1\Sigma^+$};
    \node (ES) at (7.2,4) {$B^3\Pi_1$};

    \draw[-{Latex[length=4mm,width=4mm]},line width=3pt,gray] (4,1.25) -- node[above,black,sloped]{$P2$ $F1$} (3.25,3.75);

    \tikzset{snake it/.style={-{Latex[length=2mm,width=2mm]}, decorate, decoration={snake,amplitude=2pt,pre length=2pt,post length=3pt}}}
    \draw[snake it] (2,3.75) -- node[left,xshift=-1pt]{\footnotesize0.484} (0,.25);
    \node[above=3 of J0] {\normalsize \textbf{a)}};
    \draw[snake it] (2.75,3.75) -- node[left,xshift=3pt,yshift=-10pt]{\footnotesize0.516} (3.5,1.25);

    \tikzset{biarrow/.style={{Latex[length=1.5mm,width=1.5mm]}-{Latex[length=1.5mm,width=1.5mm]}}}
    \draw[biarrow] (p1) to[bend right] node[below,sloped]{\small26.7~GHz} (p2);
    \draw[biarrow] (p2) to[bend right] node[below,sloped]{\small40.0~GHz} (p3);

\end{tikzpicture}
	\end{minipage}
	\hfill
	\begin{minipage}[c]{0.49\textwidth}
		\begin{tikzpicture}[scale=4]
    \def\len{.05}

    \def\pos{-18}
    \def\lenmark{.025}
    \def\xoffset{.18}

    \def\xa{-4.5*\len}
    \def\xb{-2.5*\len}
    \def\xc{-0.5*\len}
    \def\xd{+1.5*\len}
    \def\xe{+3.5*\len}
    \def\xf{-6.5*\len}
    \def\xg{+5.5*\len}

    \def\xexc{.8}
    \def\yexc{.6}
    \draw (\xexc+\xc, \yexc-0.003325) node (exc) {} -- (\xexc+\xc+\len, \yexc-0.003325);
    \draw (\xexc+\xb, \yexc-0.003325) -- (\xexc+\xb+\len, \yexc-0.003325);
    \draw (\xexc+\xd, \yexc-0.003325) -- (\xexc+\xd+\len, \yexc-0.003325) node (exc2) {};
    \node[align=center] (Fp1) at (\xexc+\xc, \yexc-0.003325+0.1) {$\ket{\widetilde{J^\prime}=1^-,\widetilde{F^\prime_1}=3/2,F^\prime=1}$};

    \def\yB{.2}
    \draw (\xc, \yB-.217455) -- (\xc+\len, \yB-.217455);
    \draw (\xb, \yB-.217455) -- (\xb+\len, \yB-.217455);
    \draw (\xd, \yB-.217455) -- (\xd+\len, \yB-.217455);
    \draw (\xc, \yB-.172936) -- (\xc+\len, \yB-.172936);
    \draw (\xb, \yB-.172936) -- (\xb+\len, \yB-.172936);
    \draw (\xd, \yB-.172936) -- (\xd+\len, \yB-.172936);
    \draw (\xa, \yB-.172936) -- (\xa+\len, \yB-.172936);
    \draw (\xe, \yB-.172936) -- (\xe+\len, \yB-.172936);
    \draw (\xc, \yB+.105876) -- (\xc+\len, \yB+.105876);
    \draw (\xb, \yB+.105876) -- (\xb+\len, \yB+.105876);
    \draw (\xd, \yB+.105876) -- (\xd+\len, \yB+.105876);
    \draw (\xa, \yB+.105876) -- (\xa+\len, \yB+.105876);
    \draw (\xe, \yB+.105876) -- (\xe+\len, \yB+.105876);
    \draw (\xc, \yB+.141095) -- (\xc+\len, \yB+.141095);
    \draw (\xb, \yB+.141095) -- (\xb+\len, \yB+.141095);
    \draw (\xd, \yB+.141095) -- (\xd+\len, \yB+.141095);
    \draw (\xa, \yB+.141095) -- (\xa+\len, \yB+.141095);
    \draw (\xe, \yB+.141095) -- (\xe+\len, \yB+.141095);
    \draw (\xf, \yB+.141095) -- (\xf+\len, \yB+.141095);
    \draw (\xg, \yB+.141095) -- (\xg+\len, \yB+.141095);
    \node[align=center] (J2) at (\xd-0.5*\len,\yB+.141095+.1) {$\ket{J=2^+}$};
    \node[above=0.25 of J2] {\normalsize \textbf{b)}};

    \def\xgnda{.8}
    \def\xgndb{1.3}
    \draw (\xgnda+\xc, -0.003325) -- node (gnd1) {} (\xgnda+\xc+\len, -0.003325);
    \draw (\xgnda+\xb, -0.003325) -- (\xgnda+\xb+\len, -0.003325);
    \draw (\xgnda+\xd, -0.003325) -- (\xgnda+\xd+\len, -0.003325);
    \draw (\xgndb+\xc, 0.009975)  -- node (gnd0) {} (\xgndb+\xc+\len, 0.009975);

    \def\xF{7.5*\len}
    \def\yF{.01}
    \node (F3)  at (\xF,\yB+.141095+\yF) {3};
    \node (F2a) at (\xF,\yB+.105876-\yF) {2};
    \node (F2b) at (\xF,\yB-.172936+\yF) {2};
    \node (F1)  at (\xF,\yB-.217455-\yF) {1};
    \node[above=-0.03 of F3]   {$F$};
    \def\xFf{0.10}
    \node[below=\xFf of gnd0] {$F=0$};
    \node[below=\xFf of gnd1] {$F=1$};
    \node[align=center] (J0) at (0.5*\xgnda+0.5*\xgndb,-2*\xFf) {$\ket{J=0^+}$};

   \draw [decorate,decoration={brace,amplitude=5pt,mirror,raise=5pt}] (\xF,\yB-.217455-2*\yF) -- (\xF,\yB+.105876) node [black,midway,anchor=west,xshift=5] (trans) {};
   \draw[-{Latex[length=3mm,width=3mm]},line width=2pt,gray] (trans) -- node[above,black,sloped]{$P2~F1$} (exc);
   \tikzset{snake it/.style={-{Latex[length=2mm,width=2mm]}, decorate, decoration={snake,amplitude=1.5pt,pre length=1pt,post length=2pt}}}
   \draw[snake it] (exc2) -- node[right,xshift=0pt]{0.337} (gnd0);
   \draw[snake it] (exc2) -- node[left,xshift=0pt] {0.147} (gnd1);
\end{tikzpicture}
	\end{minipage}
	\mbox{}
	\caption{Rotational cooling scheme. \parts{a} The thick solid arrow denotes a UV laser driving the $P2$ $F1$ transition; bent arrows represent microwaves, and wavy arrows indicate spontaneous emission with branching fractions as indicated.  
    The odd-parity $\tilde{J}' = 1^-$ excited state can only decay to states with $J=0^+,2^+$. Percentages under the ground-state kets are the thermal population at temperature $T_{\rm rot}=6.3\,$K, prior to rotational cooling. \parts{b} Hyperfine structure relevant to optical pumping. Decays back to $J=2^+$ are not shown. While the $P2$ $F1$ transition does not excite $\ket{J=2^+,F=3}$, this level can be coupled into the system by microwave-induced transfer to the $\ket{J=3^-,F=2,3}$ levels and subsequent stimulated emission to $\ket{J=2^+,F=1,2}$. Figure taken from \cite{Grasdijk_2021}.
    }
	\label{fig:rotational_cooling_scheme}
\end{figure*}

The dissipation required for our rotational/hyperfine cooling scheme comes from spontaneous emission in an optical transition. We perform optical pumping by exciting a transition from the ground state to the \tlfexc\, excited state, using a laser at $271.75\nm$.  The \tlfexc\, state has a natural width $\Gamma_B = 2\pi\times1.6$~MHz, so the ground state hyperfine structure is completely unresolved in the optical transition.  However, the B~$^3\Pi_1$ state has very large hyperfine splittings, so the laser addresses a single hyperfine level in the excited state. Each such state can be described in terms of quantum numbers $J^\prime$, $F_1^\prime = J^\prime + I_1$, $F^\prime = F_1^\prime + I_2$, and parity $P^\prime$.  However, the hyperfine interaction in the excited state is so strong that states with quantum numbers $J^\prime$ and $F_1^\prime$ can be strongly mixed; we indicate the approximate nature of these quantum numbers by labeling them $\tilde{J}^\prime$ and $\tilde{F}_1^\prime$.  Branching fractions for decays of these excited-state hyperfine/rotational levels are compiled in Ref.~\cite{timgren2023thesis}, calculated analytically from angular momentum couplings.

Rotational/hyperfine cooling is accomplished with a single optical pumping laser and two microwave driving fields. The $J=2$ ground state is coupled with the laser to the $\ket{\tilde{J'}^P=1^-,\,\tilde{F}_1=3/2,\,\tilde{F}=1}$ excited state, as shown in Fig.~\ref{fig:rotational_cooling_scheme}(a). We refer to this $P$-branch transition (with $\tilde{J}^\prime = J-1$) as the $P2\ F1$ transition.  Roughly half the decays from $\tilde{J}^\prime=1$ end up in the $J=0$ ground state, and nearly all of the remainder returns to $J=2$; the only loss is from branching to other vibrational states which amounts to $\lesssim1\%$ \cite{norrgard2017hyperfine,meijer2020lambda,hunter2012prospects}. 
Simultaneous with the laser excitation, resonant microwaves couple the $J=1\leftrightarrow 2$ and $J=2\leftrightarrow 3$ transitions. Repeated excitation-decay cycles accumulate population from $J=1,\,2,\,3$ into the $J=0$ state.

In the decay of $\ket{\tilde{J'}^P=1^-,\,\tilde{F'}_1=3/2,\,\tilde{F'}=1}$ into $\ket{J=0}$, branching fractions dictate that nearly 70\% of the time the $\ket{J=0,F=0}$ sublevel is populated.  We rely only on this effect to enhance the population of the $F=0$ level relative to $F=1$ as shown in Fig.~\ref{fig:rotational_cooling_scheme}(b).
(Note that in thermal equilibrium, the $F=1$ level has three times the population of $F=0$.)
With this scheme and the known rotational temperature, a maximum gain of $\maxgain$ in the $\ket{J=0, F=0}$ population can be expected for a full depletion of the $J=1,2$, and $3$ rotational ground states.

The particular hyperfine structure of TlF adds substantial complexity to the hyperfine/rotational cooling process.  Exciting a single excited-state hyperfine level with the laser, while also coupling many unresolved hyperfine states in the ground state rotational manifold, leads to a low excitation and pumping rate due to the formation of long-lived coherent dark states \cite{BerkelandBoshier2002}. In our scheme, we work to rapidly destabilize the dark states by switching polarizations of both microwave excitation fields \cite{ShumanRadiative,YeMicrowaveRemix} and by ensuring that no pair of the three excitation fields (two microwave plus one laser) are either parallel or perpendicular to each other. The photon scattering rate on the laser-driven transition, $\Gamma_\text{sc}$, is bounded by $\Gamma_\text{sc}\lesssim \Gamma_\text{B}\cdot n_e/(n_g+n_e)$ \cite{Tarbutt_2013}, where $n_e=3$ is the number of excited state sublevels and $n_g=60$ is the number of simultaneously coupled ground-state sub-levels.
This reduction in scattering rate means that substantial interaction time is needed to achieve efficient optical pumping. To accomplish this while maintaining sufficient laser intensity to maximize the excited-state population, we send multiple passes of the laser beam through the molecular beam.

In \CENTREX, following the rotational/hyperfine cooling region, an electrostatic quadrupole lens is used to collimate the molecular beam. The lens accepts transverse velocities $|v_{\perp}| < v_{\perp}^{\rm max} = 2~\vms$.  Hence, a key requirement for the rotational/hyperfine cooling is that it be effective over this full range of transverse velocities.  This corresponds to a range of Doppler shifts, $\pm \delta_D^{\rm max}$, that is much larger than the natural width of the transition:  $\delta_D^{\rm max} \approx 4.7\,\Gamma_\text{B}$.  Therefore, substantial spectral broadening of the optical pumping light is necessary.

\section{Experimental Setup}
\label{sec:experimental_overview}

\begin{figure}
    \centering
    \includegraphics[width = 0.47\textwidth]{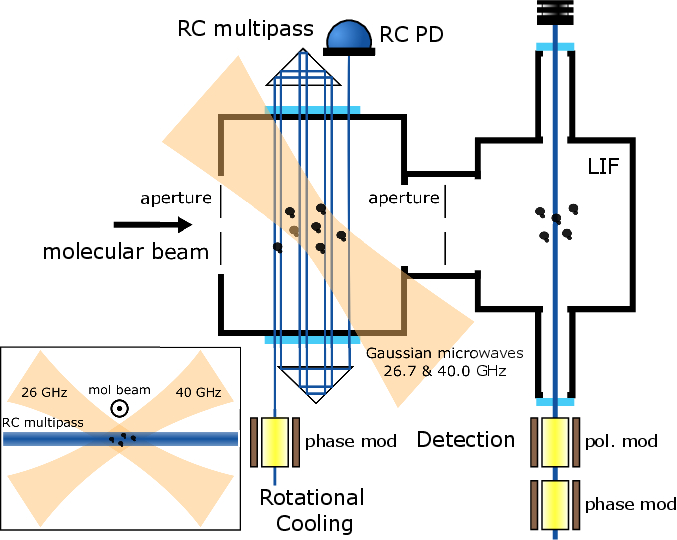}
    \caption{Schematic overview of the rotational cooling (RC) gain measurement. Two circular apertures with $8\,$mm diameter constrict the transverse velocity and spread of the molecular beam. A $271.75\nm$ laser at the \PtwoFone\,transition is phase modulated and passed 13 times through the rotational cooling region, where it intersects with two focused Gaussian microwave beams. See the inset for a frontal view of the intersecting beams. The populations are read out with laser-induced fluorescence (LIF) captured in a photomultiplier tube (PMT) from a single phase-modulated laser beam at $271.75\nm$. A photodiode (RC PD) monitors the transmitted RC light after the multi-pass.}
    \label{fig:schematic_overview}
\end{figure}

\subsection{Overview}
A schematic of the setup is shown in Figure \ref{fig:schematic_overview}. The TlF cryogenic beam, traveling horizontally with mean forward velocity $\bar{v}_f = 184$\,m/s, enters a chamber where the rotational cooling takes place, at a distance of approximately $40$~cm from the beam source and $60$~cm from the cell exit.  Here, the molecular beam is crossed simultaneously by a multi-passed laser beam and two focused, free-space microwave beams, which all serve to perform the rotational/hyperfine cooling. These beams are all nominally orthogonal to the molecules' trajectories and at $33^\circ$ angles w.r.t. each other; they are shaped so as to cover the entire vertical transverse extent of the collimated molecular beam which is a circle of $8$~mm diameter. 

The polarization of each microwave beam and the laser beam can be alternated between two orthogonal linear directions. Polarization switching makes it possible to address all hyperfine/Zeeman sublevels in the rotational state manifolds.  If the switching is sufficiently rapid (at angular frequency $\omega$ such that $\omega \gtrsim \Omega_\mu\,\text{or}\,\Gamma_{\rm B}$, where $\Omega_\mu$ is the maximum Rabi frequency of the driven microwave transitions), it also can disrupt formation of dark states \cite{BerkelandBoshier2002} that could otherwise slow the optical pumping rate.  

After this interaction region, the molecules travel downstream $\sim40$~cm to a detection region.  Here a probe laser, again propagating orthogonal to the molecular trajectories, is tuned to excite molecules in selected sublevels (see below). A photomultiplier, placed orthogonally to both the molecule and probe laser beams, detects the laser-induced $B-X$ fluorescence.

\subsection{Rotational cooling laser}
\label{sec:rot_cooling_laser}
The \PtwoFone\, optical transition \cite{norrgard2017hyperfine,meijer2020lambda} is at wavelength $271.75\nm$. The ultraviolet (UV) laser light is generated from frequency quadrupled $1087\nm$ light, using two successive second-harmonic generation (SHG) stages. A fiber seed laser is fiber-amplified to $1.5$ W. 
Resonant-cavity SHG generates $700$~mW of $543.5\nm$ light, which is fiber coupled and subsequently frequency doubled with another resonant-cavity SHG setup to produce $\sim90$~mW of $271.75\nm$ light. The resulting UV laser beam is then electro-optically phase modulated at angular frequency $\omega = \Gamma_\text{B}$ with modulation parameter $\beta = 3.8$, to generate a spectral pattern containing substantial power in the carrier and all sidebands up to the 5$^{\rm th}$ order; this covers the entire range of Doppler broadening up to $\pm \delta_D^{\rm max}$. An additional electro-optical modulator was used to rapidly switch the laser polarization for some measurements.

A pair of cylindrical lenses vertically expands the UV laser beam to a $1/e^2$ diameter of $1$~cm, while its horizontal diameter is $2$~mm. Two identical right-angle prisms, placed horizontally on either side of the molecular beam at the same height, are oriented with their $\sqrt{2}\,$-inch long hypotenuses parallel to the molecular beam forward velocity. Their positions are offset along the direction of the molecular beam by $8$~mm, so that a beam input on the uncovered edge of one prism makes 13 passes through the molecular beam before exiting the open edge of the other prism.  This multipass geometry of the laser beam extends the interaction time with the molecules to $\sim210\us$ while maintaining a high intensity as needed for efficient optical pumping. Losses in the optical path cause the intensity of the laser to reduce by a factor of $\sim8$ from the first to the last pass. The peak intensities of the multipassed beam are approximately $1000\mWcms$ for the first pass and $140\mWcms$ for the last pass.


\subsection{Microwave generation and delivery}

\begin{figure}
    \centering
    \includegraphics[width = 0.49\textwidth]{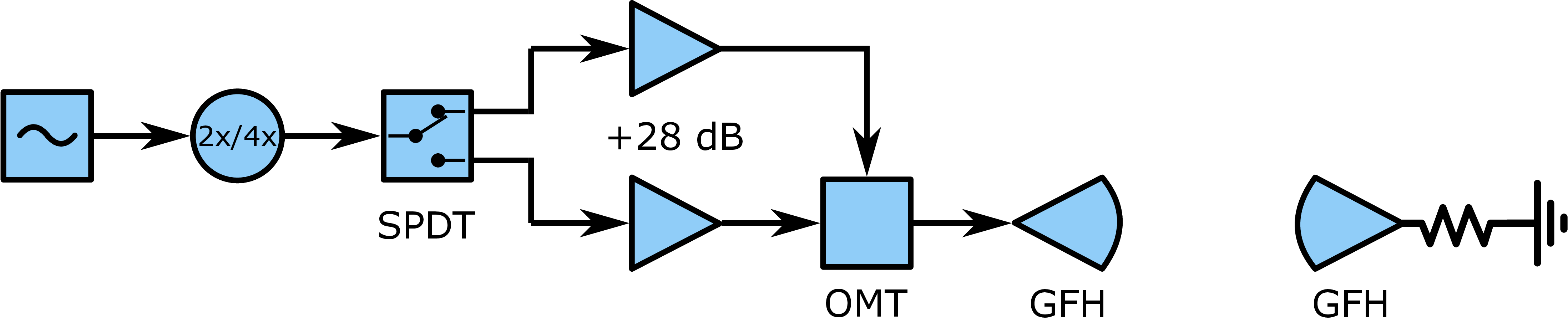}
    \caption{Schematic overview of the frequency doubled or quadrupled microwave generation system. Identical Gaussian focusing horn (GFH) lens antennas capture and terminate the microwave beams after they traverse the interaction region, to minimize reflections. The linear polarization of the microwave beams is alternated with a single-pole double-throw (SPDT) switch that directs the input microwaves through one of two circuit branches; each branch is amplified, then fed to one of the two orthogonally-polarized inputs of an orthomode transducer (OMT) whose circular-waveguide output feeds the transmitting GFH antenna.}
    \label{fig:microwave_schematic}
\end{figure}

 The rotational ground states, separated by several tens of $\GHz$, are coupled via microwave electric fields. The $J=1\leftrightarrow J=2$ transition has a resonant frequency of $26.6\GHz$, and the $J=2\leftrightarrow J=3$  transition is at $40.0\GHz$. A schematic of the microwave systems is given in Fig.~\ref{fig:microwave_schematic}.  A source signal at $13.3~(10.0)\GHz$ from a synthesizer is frequency doubled (quadrupled) to produce $26.6~(40.0) \GHz$. These microwaves are amplified, then delivered to the interaction region using cylindrically symmetric spot-focusing lens antennas, fed via circular waveguides.  These provide free-space microwave beams, focused with $-3$ dB diameters of 2.54 cm at the position of the molecules, corresponding to a transit time broadening of approximately $3.2\kHz$, given a forward velocity $\bar{v}_f=184\vms$. The microwaves enter and exit the vacuum chamber via fused silica windows. The maximum intensity at the molecules' position is $\sim68$~mW/cm$^2$, corresponding to a Rabi rate $\Omega \approx 2\pi\times 3\MHz$. The polarization of the microwave beams is rapidly switched between two orthogonal linear polarizations as shown in Fig.~\ref{fig:microwave_schematic} at a frequency of $1\MHz$.

\subsection{Detection laser}
Like the \PtwoFone\, rotational cooling transition, the \RzeroFone\, and \RzeroFtwo\, detection transitions are also at wavelength $271.75\nm$ (see below for more details). The UV laser is generated equivalently to the rotational cooling laser (Sec. \ref{sec:rot_cooling_laser}), producing $35\mW$ and expanded to a $1/e^2$ diameter of $6.5\mm$, which results in a peak intensity of $240\mW/\cm^2$. Similarly to the rotational cooling laser, this UV laser beam is electro-optically phase modulated at angular frequency $\omega=\sim\Gamma_B$ with modulation parameter $\beta=3.8$ to cover the Doppler broadening. For \RzeroFone\, an additional EOM was used to modulate the polarization at angular frequency $\omega=\sim\Gamma_B$ to destabilize dark states. The addition of polarization modulation alters the phase modulation spectrum. A small frequency difference $\Delta f$ between the phase and polarization modulation signals causes the sideband amplitudes to vary in time, asymmetrically between orthogonal polarizations, with a characteristic timescale of $1/\Delta f$. Here, $\Delta f = 65\kHz$, and molecules traversing the laser beam experience approximately two complete cycles of this polarization-dependent sideband variation. See Appendix~\ref{sec:AppendixPhaseModPolMod} for further details.

\subsection{Gain measurement schemes}
\label{sec:detection}

\begin{figure}
    \centering
    \begin{tikzpicture}[scale=4]

    \definecolor{darkgray176}{RGB}{176,176,176}
    \definecolor{darkorange25512714}{RGB}{255,127,14}
    \definecolor{lightgray204}{RGB}{204,204,204}
    \definecolor{steelblue31119180}{RGB}{31,119,180}
    \definecolor{forestgreen4416044}{RGB}{44,160,44}
    
    \def\len{.2}
    \def\spacing{1.5}
    
    \def\linewidth{2}
    
    \def\pos{-18}
    \def\lenmark{.025}
    \def\xoffset{.18}

    \def\xa{-3*\spacing*\len}
    \def\xb{-2*\spacing*\len}
    \def\xc{-1*\spacing*\len}
    \def\xd{0*\spacing*\len}
    \def\xe{1*\spacing*\len}
    \def\xf{2*\spacing*\len}
    \def\xg{3*\spacing*\len}

    \tikzset{custom_pattern/.style={dash pattern=on 3pt off 1pt}}

    \def\xexc1{0}
    \def\yexc1{.4}
    \draw[custom_pattern, line width=\linewidth pt, darkorange25512714] (\xexc1+\xb, \yexc1) -- node (F1emFnone) {} (\xexc1+\xb+\len, \yexc1);
    \draw[custom_pattern, line width=\linewidth pt, darkorange25512714] (\xexc1+\xc, \yexc1) -- node (F1emFzero) {} (\xexc1+\xc+\len, \yexc1);
    \draw[custom_pattern, line width=\linewidth pt, darkorange25512714] (\xexc1+\xd, \yexc1) -- node (F1emFpone) {} (\xexc1+\xd+\len, \yexc1);
    
    \def\xexc{0}
    \def\yexc{.6}
    \draw[line width=\linewidth pt, steelblue31119180] (\xexc+\xa, \yexc) -- node (emFntwo) {} (\xexc+\xa+\len, \yexc);
    \draw[line width=\linewidth pt, steelblue31119180] (\xexc+\xb, \yexc) -- node (emFnone) {} (\xexc+\xb+\len, \yexc);
    \draw[line width=\linewidth pt, steelblue31119180] (\xexc+\xc, \yexc) -- node (emFzero) {} (\xexc+\xc+\len, \yexc);
    \draw[line width=\linewidth pt, steelblue31119180] (\xexc+\xd, \yexc) -- node (emFpone) {} (\xexc+\xd+\len, \yexc);
    \draw[line width=\linewidth pt, steelblue31119180] (\xexc+\xe, \yexc) -- node (emFptwo) {} (\xexc+\xe+\len, \yexc);

    \def\xFzero{0}
    \def\xFone{0}
    \def\yFzero{0}
    \def\yFone{-0.1}
    \draw[line width=\linewidth pt] (\xFone+\xb, \yFone) -- node (FonemFnone) {} (\xFone+\xb+\len, \yFone);
    \draw[line width=\linewidth pt] (\xFone+\xc, \yFone) -- node (FonemFzero) {} (\xFone+\xc+\len, \yFone);
    \draw[line width=\linewidth pt] (\xFone+\xd, \yFone) -- node (FonemFpone) {} (\xFone+\xd+\len, \yFone);
    \draw[line width=\linewidth pt] (\xFzero+\xc, \yFzero)  -- node (FzeromFzero) {} (\xFzero+\xc+\len, \yFzero);

    \def\xF{7.5*\len}
    \def\yF{.2}
    \def\xFf{2*5*\spacing*\len}
    \def\xFone{2*6*\spacing*\len}
    \def\xJ{2*7*\spacing*\len}
    \def\yFf{3.0};
    \node[right=\xFf of FzeromFzero] {$0$};
    \node[right=\xFf of FonemFzero] {$1$};
    \node[right=\xFf of F1emFzero] {$1$};
    \node[right=\xFf of emFzero] {$2$};
    \coordinate (Midpoint) at ($(FonemFzero)!0.5!(FzeromFzero)$);
    \node[right=1.04*\xFone of FonemFzero] at (Midpoint) {$\sfrac{1}{2}$};
    
    \node[right=\xFone of F1emFzero] {$\sfrac{1}{2}$};
    \node[right=\xFone of emFzero] {$\sfrac{3}{2}$};
    \node[right=\xJ*1.04 of emFzero] at ($ (F1emFzero)!.5!(emFzero)$) {$1^-$};
    \node[right=\xJ*1.04 of FonemFzero] at ($ (FonemFzero)!.5!(FzeromFzero)$) {$0^+$};

    \node[below=\yFf of emFntwo] (mntwo) {$-2$};
    \node[below=\yFf of emFnone] {$-1$};
    \node[below=\yFf of emFzero] {$0$};
    \node[below=\yFf of emFpone] {$+1$};
    \node[below=\yFf of emFptwo] {$+2$};

    \node[above right=\yF and \xFf of emFzero] {$F$};
    \node[above right=\yF*0.85 and \xFone of emFzero] {$F_1$};
    \node[above right=\yF*0.85 and \xJ of emFzero] {$J^P$};
    \node[left=\spacing*\len*0 of mntwo] {$m_F$}; 

    \node[draw=steelblue31119180, line width=\linewidth pt, rounded corners=2pt, minimum width=100pt, minimum height=8pt, text centered] 
    at (FonemFzero.center) (boxFone) {};

    \node[custom_pattern, draw=darkorange25512714, line width=\linewidth pt, rounded corners=2pt, minimum width=110pt, minimum height=22pt, text centered] 
    at ($($(FonemFzero.center)!.5!(FzeromFzero.center)$) + (0, -0.4 pt)$) (boxJ0) {};

    \def\shift{\len/5}
    \coordinate (start) at ($(FonemFzero) + (-\shift, +1pt)$);
    \coordinate (end) at ($(emFzero) + (-\shift, -1pt)$);
    \draw[decorate, decoration={snake, amplitude=0.5mm, segment length=5mm}, 
              steelblue31119180, -{Stealth[scale=1.75]}, line width=\linewidth/2] 
            (start) -- (end);

    \coordinate (start) at ($(FzeromFzero) + (+\shift, +1pt)$);
    \coordinate (end) at ($(F1emFzero) + (+\shift, -1pt)$);
    \draw[decorate, decoration={snake, amplitude=0.5mm, segment length=5mm}, 
              darkorange25512714, -{Stealth[scale=1.75]}, line width=\linewidth/2] 
            (start) -- (end);
    
\end{tikzpicture}
    \caption{Detection transitions used to calculate the gain. The \RzeroFtwo\ transition (blue, solid) has no dark states and easily saturates. The \RzeroFone\ transition (orange, dashed) has dark states causing an imbalance in the relative amount of photons coming from the $F=0$ and $F=1$ ground state levels.}
    \label{fig:detection_transitions}
\end{figure}

The figure of merit for the rotational/hyperfine cooling is the gain in population of the $\ket{J=0, F=0}$ state.  It is difficult to directly probe the population of this state. The optical transition linewidth is much larger than the hyperfine splitting in the ground state, and there is no optical transition where selection rules ensure excitation only of the $\ket{J=0, F=0}$ state; hence, every probe of the $J=0$ state inevitably also addresses the $\ket{J=0, F=1}$ state.  

To extract information about the population gain in the $\ket{J=0, F=0}$ state, we use several schemes.  These are in principle equivalent, but sensitive to different systematic errors and hence provide a useful cross-check of the indirect probes. These schemes use two optical probe transitions, which we denote as \RzeroFtwo\ and \RzeroFone\, as shown in Fig.~\ref{fig:detection_transitions}, to determine population in various sublevels of the $J=0$ state. The \RzeroFtwo\ transition excites molecules only from the $F=1$ hyperfine level of the $J=0^+$ ground state, to the excited state with $J^\prime = 1^-, F1^\prime=3/2, F^\prime = 2$.  All Zeeman sublevels ($m_F=0,\pm 1$) are excited with any polarization of light.  We refer to the fluorescence signal when probing on this transition as $S_{1}$.  The \RzeroFone\ transition excites from the unresolved $F=0$ and $F=1$ hyperfine levels of the $J=0^+$ ground state, to the excited state with $\tilde{J^\prime} = 1^-, \tilde{F1^\prime}=1/2, F^\prime = 1$.  With the probe light rapidly modulated between orthogonal linear polarizations, this transition can excite population from all Zeeman sublevels of both the $F=0$ and $F=1$ levels. We refer to the fluorescence signal when probing on this transition as $S_{0+1}$. 

With these signals, we use the following schemes to deduce the gain in the $\ket{J=0, F=1}$ state population:

 \begin{itemize}
 
     \item Scheme 1: \RzeroFtwo\ branching fraction method.  Here we use $S_1$ to detect the total population in the $\ket{J=0,F=1}$ level, with rotational cooling on ($S_1^{\rm on}$) and off ($S_1^{\rm off}$). Then the gain in the population of the $\ket{J=0,F=1}$ level, $G_1$, is 
     \begin{equation}
        \label{eq:branchingR0F2}
         G_1 \equiv \frac{S_1^{\rm on}}{S_1^{\rm off}}.
     \end{equation}
     If the initial population in the $\ket{J=0,F=1}$ level is $\rho_1$, the change in its population due to rotational cooling, $\Delta\rho_1$, is
     \begin{equation}
         \Delta\rho_1 = (G_1-1)\rho_1 = 3(G_1-1)\rho_0,
     \end{equation}
     where we use the fact that the initial (thermal) population in the $\ket{J=0,F=0}$ level, $\rho_0$, is 1/3 of $\rho_1$.  The change in population of the $\ket{J=0,F=0}$ level, $\Delta\rho_0$, is given by 
     \begin{equation}
         \Delta \rho_{0} = \frac{\brzero}{\brone}\Delta\rho_1,
     \end{equation}
    where $\brzero$ ($\brone$) is the branching fraction for decay from the $J^\prime = 1^-, F1^\prime = 3/2, F' = 1$ excited state of the rotational cooling transition, into the $J=0$ $F=0$ $(F=1)$ level.  
    
    Finally, the gain in population of the $\ket{J=0,F=0}$ level, $G_0$, is then given by:
    \begin{equation}
        \label{eq:gain_branching_R0F2}
         G_{0}^{(1)} = \frac{\Delta \rho_{0}}{\rho_{0}} + 1 = 3 \frac{\brzero}{\brone} \left(G_{1} - 1\right) + 1,
    \end{equation}
    where the superscript refers to the number of the scheme.

    \item Scheme 2: \RzeroFone\ branching fraction method. Using analogous logic, we find 
    \begin{equation}
         G_{0}^{(2)} = \frac{(4G_{0+1}-3)\frac{\brzero}{\brone} + 1}{1 + \frac{\brzero}{\brone}},
         \label{eq:branchingR0F1_0}
    \end{equation}
    where, as before,
    \begin{equation}
        \label{eq:branchingR0F1}
        G_{0+1} \equiv \frac{S_{0+1}^{\rm on}}{S_{0+1}^{\rm off}}.
    \end{equation}
    Further details are provided in Appendix \ref{sec:AppendixDetectionGain}.
    
    \item Scheme 3: Differential method.  This method uses information from both probe transitions, without any need for knowledge of the branching fractions. Here,
        \begin{equation}
        \label{eq:gain_differential}
         G_{0}^{(3)} = 4G_{0+1}-3G_1.
    \end{equation}
    Additional details are given in Appendix \ref{sec:AppendixDetectionGain}.


 \end{itemize}

From both simulations and measurements, we have found that, for the available laser power and interaction time, the \RzeroFone\, transition does not result in the maximum possible number of photons scattered per molecule. We attribute this fact to the existence of dark states in this transition, despite our effort to destabilize these states by rapid polarization switching \cite{BerkelandBoshier2002}.
This results in small discrepancies between the actual gain and the gain determined with schemes that use \RzeroFone\ probe transition, i.e., our schemes 2 and 3.  By contrast, the \RzeroFtwo\, transition has no dark states; as such, our scheme 1 provides a more reliable measure of the gain from rotational/hyperfine cooling. 
The discrepancies in schemes 2 and 3 vary with the relative excitation efficiency for the $J=0\,,F=1$ level ($\epsilon_1$) versus that for the $J=0\,,F=0$ level ($\epsilon_0$): the larger the difference, the larger the apparent gain when determined using schemes 2 and 3. Figure~\ref{fig:gain_change} displays this behavior. Additional details are provided in Appendix \ref{sec:AppendixGainIncrease}.

\begin{figure}
    \centering
    \renewcommand\sffamily{}
\begin{tikzpicture}

\definecolor{darkgray176}{RGB}{176,176,176}
\definecolor{darkorange25512714}{RGB}{255,127,14}
\definecolor{lightgray204}{RGB}{204,204,204}
\definecolor{steelblue31119180}{RGB}{31,119,180}

\begin{axis}[
legend cell align={left},
legend style={fill opacity=0.8, draw opacity=1, text opacity=1, draw=lightgray204},
tick align=outside,
tick pos=left,
x grid style={darkgray176},
xlabel={\(\displaystyle \epsilon_1/\epsilon_0\)},
xmajorgrids,
xmin=0.67, xmax=1.33,
xtick style={color=black},
y grid style={darkgray176},
ylabel={apparent / actual gain},
ymajorgrids,
ymin=0.831639978717275, ymax=1.25494517508527,
ytick style={color=black}
]
\addplot [very thick, steelblue31119180]
table {%
0.7 1.23570402979582
0.712 1.22367831398991
0.724 1.21192556525147
0.736 1.20043659391614
0.748 1.18920261824547
0.76 1.17821524204074
0.772 1.167466433715
0.784 1.15694850671368
0.796 1.14665410118347
0.808 1.13657616679758
0.82 1.12670794665324
0.832 1.11704296216406
0.844 1.10757499887623
0.856 1.09829809314355
0.868 1.08920651960086
0.88 1.08029477938099
0.892 1.07155758902397
0.904 1.06298987003164
0.916 1.05458673902422
0.928 1.0463434984588
0.94 1.03825562787262
0.952 1.03031877561689
0.964 1.02252875104934
0.976 1.01488151715615
0.988 1.00737318357585
1 1
1.012 0.992758349926987
1.024 0.98564474474721
1.036 0.97865581813923
1.048 0.97178832075803
1.06 0.96503911519775
1.072 0.958405171212503
1.084 0.95188356117997
1.096 0.945471455793505
1.108 0.939166119969442
1.12 0.932964908957153
1.132 0.926865264640243
1.144 0.920864712018012
1.156 0.914960855857015
1.168 0.909151377503211
1.18 0.903434031845767
1.192 0.897806644424192
1.204 0.89226710867095
1.216 0.886813383282216
1.228 0.881443489709876
1.24 0.8761555097683
1.252 0.87094758334981
1.264 0.865817906243117
1.276 0.860764728049365
1.288 0.855786350190718
1.3 0.850881124006729
};
\addlegendentry{differential}
\addplot [dashed, very thick, darkorange25512714]
table {%
0.7 1.16411623562229
0.712 1.15574295829463
0.724 1.14755974274741
0.736 1.13956019039202
0.748 1.13173818667092
0.76 1.12408788547051
0.772 1.11660369454949
0.784 1.10928026190601
0.796 1.10211246301411
0.808 1.09509538886525
0.82 1.08822433475649
0.832 1.08149478977125
0.844 1.07490242690349
0.856 1.0684430937797
0.868 1.06211280393696
0.88 1.05590772861858
0.892 1.04982418905182
0.904 1.04385864917492
0.916 1.03800770878339
0.928 1.03226809706739
0.94 1.02663666651461
0.952 1.02111038715473
0.964 1.0156863411232
0.976 1.01036171752401
0.988 1.00513380757244
1 1
1.012 0.994957776705361
1.024 0.99000470863597
1.036 0.985138451886199
1.048 0.980356743998876
1.06 0.975657400457937
1.072 0.971038311360772
1.084 0.966497438259607
1.096 0.962032811162007
1.108 0.957642525681203
1.12 0.953324740327605
1.132 0.949077673933392
1.144 0.944899603202624
1.156 0.940788860379781
1.168 0.936743831030129
1.18 0.932762951925668
1.192 0.928844709030894
1.204 0.924987635582872
1.216 0.921190310260551
1.228 0.917451355438488
1.24 0.91376943552049
1.252 0.910143255348938
1.264 0.906571558685807
1.276 0.903053126761645
1.288 0.899586776888992
1.3 0.896171361136917
};
\addlegendentry{br. $F'=1$}
\end{axis}

\end{tikzpicture}
    \vspace*{-10pt}
    \caption{Dependence of the apparent gain found using the differential method (Scheme 3, Eq.~\ref{eq:gain_differential}) and the \RzeroFone\ branching method (Scheme 2, Eq.~\ref{eq:branchingR0F1_0}) on the ratio of detection efficiencies for the $J=0, F=1$ state, $\epsilon_1$, and for the $J=0, F=0$ state, $\epsilon_0$. The $F=0,\,m_F=0$ level has on average a smaller contribution to the dark states under different polarizations than the $F=1$ manifold, resulting in $\sfrac{\epsilon_1}{\epsilon_0}<1$.}
 \label{fig:gain_change}
 \end{figure}

\section{Results}
\label{sec:results}

\begin{figure}
    \centering
    \renewcommand\sffamily{}
    \input{gain_J0_F0_3_method.tex}
    \vspace*{-10pt}
    \caption{Gain in the $J=0\,,F=0$ population plotted against the detuning from the central \PtwoFone\, line, with a gain of $20.1(4)$ for the \RzeroFtwo\, branching method (scheme 1), a gain of $22.1(4)$ for the \RzeroFone\, branching method (scheme 2), and a gain of $22.9(6)$ for the differential method (scheme 3). The gains are calculated by averaging over the grey-shaded area, which corresponds to a $\pm 2 \vms$ detuning—the same transverse velocity acceptance range as the electrostatic lens. Both the detection and the rotational cooling laser are phase modulated for these measurements.}
    \label{fig:gain_J0}
\end{figure}
Our primary results are shown in Figure \ref{fig:gain_J0}.  Given the drawbacks of employing the \RzeroFone\, transition in detecting the rotational cooling gain, as demonstrated in  Sec.~\ref{sec:detection}, we only use the result Scheme 1 (based on the branching fractions of the \RzeroFtwo\, transition) for our quantitative conclusions (Eq.~\ref{eq:gain_branching_R0F2}).

Figure~\ref{fig:gain_J0} shows the measured gain as a function of laser detuning for the three different methods. Over the central region of transverse velocities, with $|v_\perp| < v_\perp^{\rm max} \pm 2$ m/s, the gain is independent of $v_\perp$, with average value $G_0^{(1)} = 20.1(4)$.

The method of Scheme 2 (Eq.~\ref{eq:branchingR0F1_0}) resulted in a measured gain of $G_0^{(2)} = 22.1(4)$ and the method of Scheme 3 (Eq.~\ref{eq:gain_differential}) resulted in a measured gain of $G_0^{(3)} = 22.9(6)$. As expected, due to the dark states, this gain differs from the gain of Scheme 1 ($G_0^{(1)}$), and the discrepancy is larger for Scheme 3.

We also monitored the depletion of the various levels excited in the rotational cooling process, and calculated the expected gain based on the degree of depletion. The results (depletion of $85\%, 79\%$, and $81\%$, respectively, for the $J=1, 2$, and $3$ states) corresponded to an expected gain of $G_0 \approx 24$, broadly consistent with the result of Scheme 1. However, we found evidence that the population of the initially depleted states was modified by the presence of scattered microwaves in the region between the rotational cooling laser beams and the detection laser beam. Hence, we use this information only for qualitative confirmation of our results.  Appendix \ref{sec:AppendixGainFromDepletion} contains further details.

\section{Conclusion}
\label{sec:conclusion}

We have demonstrated a rotational cooling gain of $G_0 = 20.1(4)$ in the population of the $\ket{J=0,F=0}$ state of TlF molecules, sufficient for achieving the projected statistical sensitivity of CeNTREX~\cite{Grasdijk_2021}. 

From simulations, we expect that increasing the power of the rotational cooling laser to above $\sim500\mW$ should result in full depletion of the $J=1,2$, and $3$ states over the relevant velocity range, resulting in a gain of over $25$. 
We are currently implementing a new laser system capable of achieving this power.
Another factor of $\sim1.5$ improvement in the gain could be obtained by adding a second cooling laser to pump out the $\ket{J=0,~F=1}$ hyperfine manifold via the \RzeroFtwo\ transition. This transition has no dark states and could be saturated with modest laser power. Together, these improvements could lead to a total gain of nearly 40 and a corresponding statistical improvement in the CeNTREX measurement of the $^{205}$Tl nuclear Schiff moment.

\begin{acknowledgments}
We are grateful for support from the John Templeton Foundation, the Heising-Simons Foundation, a NIST Precision Measurement Grant, NSF-MRI grants PHY-1827906, PHY-1827964, and PHY-1828097, NSF grant PHY-2110420, and the Department of Energy (DOE), Office of Science, Office of Nuclear Physics, under contract number DEAC02-06CH11357 and grant number DE-SC0024667.
\end{acknowledgments}

\appendix

\section{Detection gain}
\label{sec:AppendixDetectionGain}

Throughout this section the following notational conventions are used: $\rho_0$ ($\rho_1$) denotes the population in the $F=0$ ($F=1$) hyperfine manifold, $\rho_0^{RC}$ ($\rho_1^{RC}$) is this population after rotational cooling, $n_{\gamma,0}$ ($n_{\gamma,1}$) denotes the number of photons scattered for $F=0$ ($F=1$), and $G_0$ ($G_1$) is the population gain in the $F=0$ hyperfine manifold. The signal for a transition is denoted by $S$, where a superscript of $RC$ indicates a signal after rotational cooling.

The \RzeroFtwo\, transition addresses only the $F=1$ hyperfine level in the $J=0$ rotational manifold, and the signal is described as
\begin{equation}
    S_{R0F2} = S_1=\rho_1\cdot n_{\gamma,1}.
\end{equation}

The \RzeroFone\, transition addresses the $F=0$ and $F=1$ hyperfine levels in the $J=0$ rotational manifold, and the signal is denoted by
\begin{equation}
    S_{R0F1} = S_{0+1}=\rho_0\cdot n_{\gamma,0}+\rho_{1}\cdot n_{\gamma,1},
\end{equation}
 where, under the assumption of optical cycling to completion, it can be written as
\begin{equation}
    S_{0+1}=\left(\rho_{0} + \rho_{1}\right)\cdot n_{\gamma}.
\end{equation}

Further calculations assume that when the population is thermally distributed, it is evenly spread over the hyperfine sublevels of a single rotational manifold, hence $\rho_{1} = 3\rho_{0}$.\\

With the known branching fractions for the \PtwoFone\, transition to each of the $F=0$ and $F=1$ hyperfine manifolds, the individual transitions can be used to calculate $G_{0}$. Starting with \RzeroFtwo:\\  
\begin{equation}
    \Delta \rho_{1} = \rho_{1}^{RC} - \rho_{1} = \rho_{1}\left(G_{1}-1\right),
\end{equation}
\begin{equation}
    \Delta \rho_{0} = \frac{\brzero}{\brone}\Delta\rho_{1}.
\end{equation}
Using $3\rho_{0} = \rho_{1}$ the increase in population to $\rho_{0}$ is given by
\begin{equation}
    \Delta \rho_{0} = 3 \frac{\brzero}{\brone} \rho_{0} \left(G_{1}-1\right).
\end{equation}
Finally, the gain in $F=0$ is then given by
\begin{equation}
    \label{eq:gain_branching_R0F2_appendix}
    G_{0}^{(1)} = \frac{\rho_{0}^{RC}}{\rho_{0}} = 3 \frac{\brzero}{\brone} \left(G_{1} - 1\right) + 1,
\end{equation}
where the superscript refers to the number of the scheme.  This is the same result as in Eq. (\ref{eq:gain_branching_R0F2}).

For \RzeroFone\, the gain $G_{0}$ can also be calculated with the branching fractions.  First we invert Eq.~(\ref{eq:gain_branching_R0F2_appendix}) to get
\begin{equation}
    G_{1} = \frac{G_{0}-1}{3\frac{br_{F=0}}{br_{F=1}}} + 1.
\end{equation}
Substituting this into Eq.~(\ref{eq:gain_differential_appendix}) below, we obtain
\begin{equation}
    G_{0}^{(2)} = \frac{(4G_{0+1}-3)\frac{\brzero}{\brone} + 1}{1 + \frac{\brzero}{\brone}}.
\end{equation}

The differential method of calculating the gain uses both transitions. The gain in signal size for each transition can be written as
\begin{equation}
    G_{1} = \frac{S^{\rm on}_{1}}{S^{\rm off}_{1}} = \frac{\rho_{1}^{RC}}{3\rho_{0}},
\end{equation}
\begin{equation}
    \label{eq:gain_0p1}
    G_{0+1} = \frac{S^{\rm on}_{0+1}}{S^{\rm off}_{0+1}} = \frac{\rho_{0}^{RC}+\rho_{1}^{RC}}{4\rho_{0}}.
\end{equation}
The gain in $F=0$ is then given by
\begin{equation}
    \label{eq:gain_differential_appendix}
    G_{0}^{(3)} = 4 G_{0+1} - 3 G_{1}.
\end{equation}

\section{Apparent gain increase}
\label{sec:AppendixGainIncrease}

Under incomplete optical cycling of the \RzeroFone\,detection transition, a larger apparent gain is measured. This stems from the imbalance between the detection efficiency of the $F=1$ levels and the $F=0$ level of the $J=0$ ground state manifold. The \RzeroFone\, transition has 4 ground state levels and 3 excited state levels, meaning there is always a single dark state, for any polarization. The $F=0,\,m_F=0$ level on average has a smaller contribution to the dark states under different polarizations than the $F=1$ manifold. Due to the increase in population after RC, which predominantly pumps into $F=0$, a larger fraction of the detected photons per molecule comes from $F=0$ with rotational cooling than without it. The detection methods employing \RzeroFone\, all rely on the ratio between the signal with and without RC ($S^{\rm on}_{0+1}/S^{\rm off}_{0+1}$) and this results in a larger ratio than would be expected based on true population increases.

Substituting $\rho_i \epsilon_i$ for $\rho_i$, using $g = \sfrac{\rho^{RC}_{0}}{\rho_{0}}$ and $\rho_1 = $ in Eq.~\ref{eq:gain_0p1}
For a given gain $g = \sfrac{\rho^{RC}_{0}}{\rho_{0}}$, $\bronezero = \sfrac{\brone}{\brzero}$ and detection efficiencies $\epsilon_0,~\epsilon_1$ for $F=0,~F=1$, respectively, the signal ratio for \RzeroFone\, can be described by
\begin{equation}
    \frac{S^{\rm on}_{0+1}}{S^{\rm off}_{0+1}} = \frac{g\epsilon_0 + [3+ (g-1) \bronezero] \epsilon_1}{\epsilon_0 + 3\epsilon_1},
\end{equation}
and by the signal ratio for \RzeroFtwo\, can be described by
\begin{equation}
    \frac{S^{\rm on}_{1}}{S^{\rm off}_{1}} = \frac{\left(g-1\right)br_{1/0}+3}{3}.
\end{equation}
Equation~(\ref{eq:gain_differential}) for calculating the gain from a differential measurement can then be rewritten as
\begin{equation}
    \label{eq:gain_3_measured}
    G_{0}^{(3)} = \frac{[3+\bronezero(g-1)]\sfrac{\epsilon_1}{\epsilon_0} + 4g-3+\bronezero(1-g)}{1+3\sfrac{\epsilon_1}{\epsilon_0}},
\end{equation}
and Eq.~(\ref{eq:branchingR0F1_0}) for calculating the gain from \RzeroFone\, with the branching fractions can be rewritten as
\begin{equation}
    \label{eq:gain_2_measured}
    \begin{split}
        G_{0}^{(2)} &= \frac{\left[4\bronezero\left(g-1\right)+12\right]\sfrac{\epsilon_1}{\epsilon_0}}{\left(\bronezero+1\right)\left(1+3\sfrac{\epsilon_1}{\epsilon_0}\right)} \\
        &+ \frac{\left(\bronezero-3\right)\left(1+3\sfrac{\epsilon_1}{\epsilon_0}\right) + 4g}{{\left(\bronezero+1\right)\left(1+3\sfrac{\epsilon_1}{\epsilon_0}\right)}}.
    \end{split}
\end{equation}

The $F=0,\,m_F=0$ level has on average a smaller contribution to the dark states under different polarizations than the $F=1$ manifold, such that $\sfrac{\epsilon_1}{\epsilon_0} < 1$, resulting in an increase in the measured gain for both Eqs.~\ref{eq:gain_3_measured} and~\ref{eq:gain_2_measured}.

\section{Gain from measured depletion of rotational states}
\label{sec:AppendixGainFromDepletion}

Monitoring the extent to which the population of the rotational levels $J=1,2,3$ is depleted in the rotational cooling process provides an independent scheme for determining the rotational cooling gain $G_0$.  The depletions are determined by measuring the \RoneFthree,~\RtwoFfour~and~\RthreeFfive~transitions while toggling the RC on/off.  Below, $d_i$ corresponds to the ratio between the $R(i)$ transition with RC on and off.

This method requires an independent knowledge of the rotational temperature, $T$.  We find 
\begin{equation}
    G_{0}^{(4)} = 1 + \frac{\brzero}{\brzero + \brone}\frac{d_1\rho_{1,T} + d_2\rho_{2,T} + d_3\rho_{3,T}}{\rho_{0,T}},
\end{equation}
where $\rho_{i,T}$ and $d_i$ are the thermal population and depletion factor of $J=i$, respectively.

\begin{figure}
    \centering
    \renewcommand\sffamily{}
    \input{depletions_R_branch}
    \vspace*{-10pt}
    \caption{Depletion for the \RoneFthree,\,\RtwoFfour\, and \RthreeFfive\, transitions, plotted agains detuning from the central \PtwoFone\, line. Observed depletions are $0.85(2)$, $0.79(2)$ and $0.81(2)$ for $R(1)$, $R(2)$ and $R(3)$, respectively. For $T_\mathrm{rot}=6.3(2)$\,K this corresponds to a gain of $24.1(11)$ in $J=0,\,F=0$. The gains are calculated by averaging over the grey-shaded area, which corresponds to a $\pm 2 \vms$ detuning—the same transverse velocity acceptance range as the electrostatic lens. Both the detection and the rotational cooling laser are phase modulated for these measurements.}
    \label{fig:depletions}
\end{figure}

Figure~\ref{fig:depletions} shows the measured depletions, corresponding to loss of population by factors of $0.85(2)$, $0.79(2)$ and $0.81(2)$ for $R(1)$, $R(2)$ and $R(3)$, respectively. This would correspond to a total expected gain of $G_0 = 24.1(1.1)$, assuming $T_\mathrm{rot}=6.3(2)$\,K and equal levels of depletion over all hyperfine states in a rotational manifold. As mentioned in the main text, we suspect that the modest discrepancy with the result from our primary measurement method arises because, during these measurements, microwaves leaked into the detection chamber and affected the measured depletion ratios. 

\section{Branching Fraction Calculation}

The branching fractions for decays from different hyperfine levels of the \tlfexc\, state must be known in order to compute the rotational cooling efficiency. These branching fractions are relevant for both scheme 1 and scheme 2. The calculation is nontrivial due to the exceptionally large magnetic hyperfine splitting in the \tlfexc\, state, which leads to significant mixing between states of different $J$~\cite{norrgard2017hyperfine}. To compute the branching fractions, we proceed as follows:
\begin{itemize}
\item Diagonalize the \tlfexc\, Hamiltonian to obtain the excited-state eigenstates, which are superpositions of basis states with different $J$ and $F_1$, but share the same total angular momentum $F$.
\item Compute the electric dipole matrix elements between \tlfground\, and \tlfexc\,  basis states with definite $J$.
\item Use the mixing coefficients and these matrix elements to determine the branching fractions for decay from each \tlfexc\, eigenstate.
\end{itemize}

The electric dipole matrix elements between \tlfexc\, and \tlfground\, states are given by
\begin{equation}
\begin{split}
M = &\bra{\tlfgroundmath, \Omega, I_1, I_2, J, F_1, F, m_F} d_p^{(1)} \\
&\quad \ket{\tlfexcmath, \Omega', I_1', I_2', J', F_1', F', m_F'}
\end{split}
\end{equation}
where $d_p^{(1)}$ is the spherical component $p$ of the electric dipole operator in the lab frame. Applying the Wigner-Eckart theorem to factor out the angular dependence yields
\begin{equation}
M = \braket{F_1', m_F'; 1, q}{F, m_F} M_r,
\end{equation}
where
{
\begin{equation}
M_r = \bra{\mathrm{X}, \Omega, I_1, I_2, J, F_1, F} | d^{(1)} | \ket{\mathrm{B}, \Omega', I_1', I_2', J', F_1', F'}.
\end{equation}
}

Since the electric dipole operator does not act on nuclear spin degrees of freedom, we apply the spectator theorem to decouple $F$, $F_1$, and $I_2$, followed by $F_1$, $J$, and $I_1$. Transforming to the molecule-fixed frame gives
\begin{equation}
\begin{split}
M_r &= \bra{\mathrm{X}} |d^{(1)}| \ket{\mathrm{B}}
(-1)^{F_1' + F_1 + F' + I_1 + I_2 - \Omega} \\
&\quad \times \left[(2J + 1)(2J' + 1)(2F + 1) \right. \\
&\qquad \left. (2F' + 1)(2F_1 + 1)(2F_1' + 1)\right]^{1/2} \\
&\quad \times
\begin{Bmatrix}
F_1' & F' & I_2 \\
F & F_1 & 1
\end{Bmatrix}
\begin{Bmatrix}
J' & F_1' & I_1 \\
F_1 & J & 1
\end{Bmatrix}
\sum_{q = -1}^{1}
\begin{pmatrix}
J & 1 & J' \\
-\Omega & q & \Omega'
\end{pmatrix}
\end{split}
\end{equation}

This expression suffices for our purposes, as the reduced matrix element $\bra{\mathrm{X}}|d^{(1)}|\ket{\mathrm{B}}$ is common to all decays of interest.

The branching fraction between an initial state $\ket{i}$ and a final state $\ket{f}$ is given by
\begin{equation}
bf_{i \to f}=\frac{\Gamma_{i\to f}}{\Gamma_\mathrm{tot}}
\end{equation}
where $\Gamma_{i \to f}$ is the decay rate from $\ket{i}$ to $\ket{f}$ and $\Gamma_\mathrm{tot} = \sum_{f'} \Gamma_{i \to f'}$ is the total decay rate summed over all final states~\cite{timgren2023thesis}. For an initial state with total angular momentum $F'$, the decay rate from $\ket{i}$ to $\ket{f}$ is given by
\begin{equation}
\Gamma_{i\to f} = \frac{3\omega_0^3}{3\hbar c^3} \frac{\left|\bra{i}|d|\ket{f}\right|^2}{2F'+1},
\end{equation}
assuming an electric dipole transition. Here, $\omega_0$ is taken to be approximately constant across all transitions, since they all couple to the vibrational ground state at a frequency near $1100\THz$, with rotational splittings of only several $10\GHz$.

Although the excited-state eigenstates are mixtures of components with different $J$ and $F_1$, they possess well-defined total angular momentum $F$. As a result, differences in decay rates stem solely from variations in the reduced matrix elements. Thus, we may write
\begin{equation}
\Gamma_{\text{tot}} = \sum_{f'} \Gamma_{i \to f'} = \frac{4 \omega_0^3}{3 \hbar c^3 (2F'+1)} \sum_{f'} \left| \bra{i} d \ket{f'} \right|^2,
\end{equation}
and the rotational branching fractions are given by
\begin{equation}
bf_{i\to f} = \frac{\left| \bra{i} |d| \ket{f} \right|^2}{\sum_{f'} \left| \bra{i} |d| \ket{f'} \right|^2}.
\end{equation}
The initial states are mixtures and take the form $\ket{i} = \sum_j c_j \ket{i_j}$, leading to reduced matrix elements
\begin{equation}
\bra{i}|d|\ket{f} = \sum_j c_{j}\bra{i_j}|d|\ket{f}.
\end{equation}
Inserting this into the expression for the branching fraction yields
\begin{equation}
bf_{i\to f} = \frac{\left| \sum_j c_{j}\bra{i_j}|d|\ket{f} \right|^2}{\sum_{f'} \left| \sum_j c_{j}\bra{i_j}|d|\ket{f'} \right|^2},
\end{equation}
or in terms of $M_r$:
\begin{equation}
    bf_{i\to f} = \frac{\left|\sum_j c_j M_r(i_j\to f)\right|^2}{\sum_{f'} \left|\sum_j c_j M_r(i_j\to f')\right|^2}.
\end{equation}

These branching fractions are then evaluated for the eigenstates of the Hamiltonian obtained from~\cite{meijer2020lambda}; see Table~\ref{tab:branching_fractions}.

\begin{table}
    \makebox[\columnwidth][c]{%
        \resizebox{\columnwidth}{!}{
        \begin{tabular}{llllll}
        \toprule
         & $bf_{\tilde{J}\rightarrow J=\tilde{J}-2}$ & $bf_{\tilde{J}\rightarrow J=\tilde{J}-1}$ & $bf_{\tilde{J}\rightarrow J=\tilde{J}}$ & $bf_{\tilde{J}\rightarrow J=\tilde{J}+1}$ & $bf_{\tilde{J}\rightarrow J=\tilde{J}+2}$ \\
        \midrule
        $\ket{\tilde{J} = 1, \tilde{F_1} = 1/2, F = 0}$ &  & 0.6667 & 1.0000 & 0.3333 &  \\ \\
        $\ket{\tilde{J} = 1, \tilde{F_1} = 1/2, F = 1}$ &  & 0.6665 & 0.9999 & 0.3335 & 0.0001 \\
        $\ket{\tilde{J} = 1, \tilde{F_1} = 3/2, F = 1}$ &  & 0.4841 & 0.8907 & 0.5159 & 0.1093 \\
        $\ket{\tilde{J} = 2, \tilde{F_1} = 3/2, F = 1}$ & 0.1827 & 0.7095 & 0.8173 & 0.2905 &  \\ \\
        $\ket{\tilde{J} = 1, \tilde{F_1} = 3/2, F = 2}$ &  & 0.4797 & 0.8880 & 0.5203 & 0.1120 \\
        $\ket{\tilde{J} = 2, \tilde{F_1} = 3/2, F = 2}$ & 0.1869 & 0.7119 & 0.8130 & 0.2881 & 0.00005 \\
        $\ket{\tilde{J} = 2, \tilde{F_1} = 5/2, F = 2}$ & 0.00007 & 0.5250 & 0.9465 & 0.4750 & 0.0534 \\
        $\ket{\tilde{J} = 3, \tilde{F_1} = 5/2, F = 2}$ & 0.0751 & 0.6249 & 0.9249 & 0.3751 &  \\ \\
        $\ket{\tilde{J} = 2, \tilde{F_1} = 5/2, F = 3}$ &  & 0.5235 & 0.9456 & 0.4765 & 0.0544 \\
        $\ket{\tilde{J} = 3, \tilde{F_1} = 5/2, F = 3}$ & 0.0764 & 0.6258 & 0.9235 & 0.3742 & 0.00003 \\
        $\ket{\tilde{J} = 3, \tilde{F_1} = 7/2, F = 3}$ & 0.00004 & 0.5308 & 0.9685 & 0.4692 & 0.0314 \\
        $\ket{\tilde{J} = 4, \tilde{F_1} = 7/2, F = 3}$ & 0.0407 & 0.5870 & 0.9593 & 0.4130 &  \\
        \bottomrule
        \end{tabular}
    }}
    \caption{Branching fractions from the \tlfexc, state into ground rotational levels with specified $J$ in the \tlfground\, state, computed using the eigenstates of the Hamiltonian from~\cite{meijer2020lambda}. For each \tlfexc\, parity, the corresponding branching fractions are presented on the same row in alternating columns. The parity of the ground state follows $P = (-1)^J$. The values reported here differ significantly from those in~\cite{norrgard2017hyperfine}; however, after correspondence with the authors, we believe the values presented here to be correct.}
    \label{tab:branching_fractions}
\end{table}

\section{Phase modulation followed by polarization modulation}
\label{sec:AppendixPhaseModPolMod}
To destabilize dark states in the \RzeroFone\, transition, the detection laser, which is initially linearly polarized, is simultaneously modulated in both phase and polarization for certain measurements. A phase-modulating EOM is followed by a quarter-wave plate (QWP), which converts the light to circular polarization before it enters a polarization-modulating EOM. Although the two modulators are driven at nearly the same frequency, we assume identical modulation frequencies for analytical purposes, with a slowly varying phase offset between them:
\begin{equation}
    E(t) = \frac{
e^{-i\!\Bigl(\beta\sin\!(\Omega t+\varphi(t))
                 +\tfrac{\gamma}{2}\sin(\Omega t)\Bigr)}}{2}
\begin{bmatrix}
e^{\,i\gamma\sin(\Omega t)} + i\\[6pt]
-\,e^{\,i\gamma\sin(\Omega t)} + i
\end{bmatrix},
\end{equation}
where $\beta$ is the phase modulation depth, $\gamma$ is the polarization modulation depth, and $\varphi(t)$ is the time-varying phase offset, defined as $\Delta\Omega t$. 
By applying the Jacobi-Anger expansion, the sideband structure can be expressed as:
\begin{equation}
    E_{x/y}(t) = \frac{1}{2}\sum_k C_k^{(x/y)}(\beta, \gamma, t) e^{-ik\Omega t},
\end{equation}
where the sideband amplitudes $C_k^{(x/y)}(\beta, \gamma, t)$ are given by:
\begin{subequations}
\begin{align}
C_{k}^{(x)}(\beta, \gamma, t) &= \sum_m J_m(\beta)\, e^{-i m \varphi(t)} 
\notag \\
&\quad \times \biggl[
    J_{m-k}\!\left(\tfrac{\gamma}{2}\right)
  + i\, J_{k-m}\!\left(\tfrac{\gamma}{2}\right)
\biggr], \\
C_{k}^{(y)}(\beta, \gamma, t) &= \sum_m J_m(\beta)\, e^{-i m \varphi(t)} 
\notag \\
&\quad \times \biggl[
   - J_{m-k}\!\left(\tfrac{\gamma}{2}\right)
  + i\, J_{k-m}\!\left(\tfrac{\gamma}{2}\right)
\biggr].
\end{align}
\end{subequations}

A constant phase offset leads to an asymmetry in the sideband amplitudes between $x$ and $y$ polarization components, as illustrated in Fig.~\ref{fig:simultaneous_phase_pol_mod_phase_offset} for several representative values of $\varphi$ in $C_k^{(x)}$. The corresponding $C_k^{(y)}$ sidebands differ from $C_k^{(x)}$ by a relative $\pi$ phase shift in the phase modulation, and are mirror symmetric about $k = 0$.

\begin{figure}
    \centering
    \renewcommand\sffamily{}
    \resizebox{\columnwidth}{!}{
\begin{tikzpicture}

\definecolor{darkgray176}{RGB}{176,176,176}

\begin{axis}[
colorbar,
colorbar style={ylabel={intensity}},
colormap/viridis,
point meta max=0.417407303742392,
point meta min=0.00215546566890029,
tick align=outside,
tick pos=left,
x grid style={darkgray176},
xlabel={sideband},
xmin=-8.5, xmax=8.5,
xtick style={color=black},
y grid style={darkgray176},
ylabel={time (\(\displaystyle 2\pi / \Delta\Omega\))},
ymin=0, ymax=1,
ytick style={color=black},
xtick={-8, -6, -4, -2, 0, 2, 4, 6, 8},
tick label style={/pgf/number format/fixed, /pgf/number format/precision=2},]
\addplot graphics [includegraphics cmd=\pgfimage,xmin=-8.5, xmax=8.5, ymin=0, ymax=1] {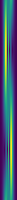};
\end{axis}

\end{tikzpicture}}
    \vspace*{-10pt}
    \caption{Time-varying sideband intensities $|C_k^{(x)}\left(\beta, \gamma, t\right)|^2$ for a phase-modulation EOM followed by a polarization-modulation EOM, both driven at the same frequency, with a time-dependent phase offset $\Delta\Omega t$ between them. The modulation depths are $\beta = 3.8$ and $\gamma = \pi/2$, respectively.}
    \label{fig:simultaneous_phase_pol_mod_phase_time_varying}
\end{figure}

When the phase offset varies in time as $\varphi(t) = \Delta\Omega t$, the sideband amplitudes oscillate between the configurations shown in Fig.~\ref{fig:simultaneous_phase_pol_mod_phase_offset}, with a characteristic timescale given by $2\pi/\Delta\Omega$, illustrated in Fig.~\ref{fig:simultaneous_phase_pol_mod_phase_time_varying} for $C_k^{(x)}\left(3.8, \pi/2, t\right)$. Again the corresponding $C_k^{(y)}$ sidebands differ from $C_k^{(x)}$ by a relative $\pi$ phase shift in the phase modulation, and are mirror symmetric about $k = 0$.


\begin{figure}
    \centering
    \renewcommand\sffamily{}
    \input{simultaneous_phase_pol_mod_phase_offset}
    \vspace*{-10pt}
    \caption{Sideband intensities $|C_k^{(x)}|^2$ for a phase-modulation EOM followed by a polarization-modulation EOM, both driven at the same frequency, with modulation depths $\beta = 3.8$ and $\gamma = \pi/2$. Curves are shown for several constant phase offsets between the phase and polarization modulation signals. Gaussian envelopes are included for illustrative purposes only.}
    \label{fig:simultaneous_phase_pol_mod_phase_offset}
\end{figure}

\bibliography{references}

\end{document}